\def\noeditingmarks{}

\documentclass[sigconf]{acmart}
\renewcommand\footnotetextcopyrightpermission[1]{} % removes footnote with conference info
\setcopyright{none}
\AtBeginDocument{\pagestyle{plain}}
\acmDOI{}
\acmISBN{}
\acmPrice{}

\usepackage[utf8]{inputenc}
\usepackage{mdframed}
\usepackage{subcaption} 
\usepackage{booktabs}  % for tables
\usepackage{dcolumn}   % for tables
\usepackage{multirow}  % for tables
\usepackage{rotating}
\usepackage{xspace}
\usepackage{hyphenat}  % supplies \hyp{}, which tells tex that it can
\usepackage{color}
\usepackage{hyperref}
\usepackage{enumerate}
\usepackage{multicol}
\usepackage{fancyvrb}
\fvset{%
fontsize=\footnotesize,
numbers=left,
xleftmargin=18pt
}
\usepackage{textcomp}
\usepackage{amsmath}
\usepackage{amsfonts}
\usepackage{wasysym}
\usepackage{tabularx}
\usepackage{pifont}
\usepackage{microtype} % for better typography
\usepackage{mathtools} % reduce vertical spacing in align environments
\usepackage{mathrsfs} % font

\usepackage{url}

\usepackage{ulem}
\usepackage{array}    % needed for 'b' argument in tabular preamble

\usepackage{float}

\usepackage[noend]{algpseudocode}
\algrenewcomment[1]{\hfill// #1}%   % don't include end-block syntax
\algtext*{Function}
\algrenewcommand{\algorithmicrequire}{\textbf{Input:}}
\algrenewcommand{\algorithmicensure}{\textbf{Output:}}
\algnewcommand{\LineComment}[1]{\State // #1}
\algrenewcommand\textproc{}% Used to be \textsc

\usepackage[noeka]{mathrmletter}
\usepackage{arydshln}
\usepackage[english]{babel}
\usepackage{blindtext}

\usepackage[font=small]{caption}
\usepackage[labelfont=small]{caption}
\usepackage[labelfont=small]{subcaption}
\usepackage{bigdelim}
\usepackage{siunitx}
\newif\ifintrouble
\newif\ifcuttext

\ifintrouble
\cuttexttrue
\else
\cuttextfalse
\fi

\usepackage{colortbl} % colorful columns and rows

\newcommand{\textred}[1]{\begingroup \color{red} #1\endgroup}
\ifx\noeditingmarks\undefined
   \newcommand{\pgwrapper}[2]{\textred{#1: #2}}
   \newcommand{\pgwrapperb}[1]{\textbf{#1}}
\else
   \newcommand{\pgwrapperb}[1]{}
   \newcommand{\pgwrapper}[2]{}
\fi

\ifx\noeditingmarks\undefined
    \newcommand{\changebars}[2]{%
    [{\em \begingroup \color{magenta} #1 \endgroup}]
    [\begingroup \color{magenta} \sout{#2} \endgroup]}
    
\else
    \newcommand{\changebars}[2]{#1}
    
\fi

\newcommand{\sys}{{Colo}\xspace}
\newcommand{\Sys}{\sys}

\newcommand{\goodcitationsize}{\fontsize{8.75}{10.5}\selectfont}
\makeatletter
\renewcommand*{\@fnsymbol}[1]{\ensuremath{\ifcase#1\or \star\or \dagger\or \ddagger\or
   \mathsection\or \mathparagraph\or \|\or **\or \dagger\dagger
   \or \ddagger\ddagger \else\@ctrerr\fi}}
\makeatother

\def\hn{\usefont{OT1}{phv}{mc}{n}\selectfont}

\newcommand{\mpfont}{\hn\scriptsize}
\newcommand{\MPworker}[2]{{\color{#1}\vrule\vrule}{\marginpar{\color{#1}\mpfont #2}}}
\ifx\noeditingmarks\undefined
    \newcommand{\MP}[1]{\MPworker{red}{#1}}
    \newcommand{\MPtg}[1]{\MPworker{red}{#1}}
    \newcommand{\MPnc}[1]{\MPworker{blue}{#1}}
    \newcommand{\MPkl}[1]{\MPworker{brown}{#1}}
\else
   \newcommand{\MP}[1]{}
    \newcommand{\MPtg}[1]{}
     \newcommand{\MPnc}[1]{}
    \newcommand{\MPkl}[1]{}
\fi
\newcommand\rmv[1]{}

\newcommand{\techReportOnly}[1]{}

\DeclarePairedDelimiterX{\norm}[1]{\lVert}{\rVert}{#1}

\usepackage{tkz-euclide}
\usepackage{tikz}

\usepackage{mathtools}
\newcommand{\cpu}{\textsc{cpu}\xspace}

\def\compactify{\itemsep0in \topsep0.5pt \parsep=0.00in \partopsep=0pt
\leftmargin2em}
\let\latexusecounter=\usecounter

\newenvironment{myenumerate}
  {\def\usecounter{\compactify\latexusecounter}
   \begin{enumerate}}
  {\end{enumerate}\let\usecounter=\latexusecounter}

\newenvironment{myenumerate2}
  {\def\usecounter{\itemsep=0ex \topsep0.7ex \parsep=1ex \partopsep=0pt
    \leftmargin1.5em\latexusecounter}
   \begin{enumerate}}
  {\end{enumerate}\let\usecounter=\latexusecounter}

\newenvironment{myitemize}%
  {\begin{list}{\labelitemi}{\itemsep3pt \topsep3pt \parsep0.00in
  \partopsep=3pt \leftmargin1em}}%
  {\end{list}}

  {\begin{list}{\labelitemi}{\itemsep6pt \topsep6pt \parsep0.00in
  \partopsep=0pt \leftmargin\parindent}}%
  {\end{list}}

  {\begin{list}{\labelitemi}{\itemsep3pt \topsep3pt \parsep0.00in
  \partopsep=0pt \leftmargin\parindent}}%
  {\end{list}}
  {\begin{list}{\labelitemi}{\itemsep0in \topsep2pt \parsep0.00in
  \partopsep=0pt \leftmargin\parindent}}%
  {\end{list}}

  {\begin{list}{\labelitemi}{\itemsep3pt \topsep3pt \parsep0.00in
  \partopsep=0pt \leftmargin1.5em}}%
  {\end{list}}

\def\emparagraph#1{\vspace{1mm}\noindent{\bf #1}}

\usepackage{amsthm}

\def\discretionaryslash{\discretionary{/}{}{/}}
{\catcode`\/\active
\gdef\URLprepare{\catcode`\/\active\let/\discretionaryslash
        \def~{\char`\~}}}%
\def\URL{\bgroup\URLprepare\realURL}%
\def\realURL#1{\tt #1\egroup}%

\begin{document}

\newcommand{\supsyml}[1]{\raisebox{4pt}{\footnotesize #1}}
\newcommand{\rstar}{\supsyml{$\ast$}\xspace}
\newcommand{\rdag}{\supsyml{$\ast\ast$}\xspace}

\date{}

\title{Making Privacy-preserving Federated Graph Analytics with Strong Guarantees Practical (for Certain Queries)}

\author{Kunlong Liu}
\affiliation{\institution{\fontsize{9.5}{11}\selectfont \textit{University of California, Santa Barbara}}\country{}}
\author{Trinabh Gupta}
\affiliation{\institution{\fontsize{9.5}{11}\selectfont \textit{University of California, Santa Barbara}}\country{}}

\begin{abstract}
    Privacy-preserving federated graph analytics is an emerging area of research.
    The goal is to run graph analytics queries over a set of devices that are organized as a graph while keeping the raw data on the devices rather than centralizing it.
    Further, no entity may learn any new information except for the final query result. For instance, a device may not learn 
    a neighbor's data.
    The state-of-the-art prior work for this problem provides privacy guarantees (but not integrity) for a broad set of queries 
    in a strong threat model where the devices can be malicious. However,  it
    imposes an impractical overhead: each device locally requires over 8.79 hours of \cpu time and 5.73~GiBs of network transfers per query.
    This paper presents Colo, a new, low-cost system
    for privacy-preserving federated graph analytics 
    that requires minutes of \cpu time and a few MiBs in network transfers, 
    for a particular subset of queries.
    At the heart of Colo is
    a new secure computation protocol that enables a device to securely and efficiently evaluate a graph query in its local neighborhood while hiding device data, edge data, and topology data.
    An implementation and evaluation of Colo shows
    that for running a variety of COVID-19 queries over a population of 1M devices,
    it requires less than 8.4 minutes of a device's \cpu time and 4.93~MiBs in network transfers---improvements of up to three orders of magnitude.
\end{abstract}
\maketitle

\section{Introduction}
\label{s:introduction}
As a motivating example, consider the following scenario between a mobile app maker of a
contact tracing application for an infectious disease like COVID-19~\cite{tracetogether,covidsafe,swisscovid}, and an influential data analyst such as the Centers for Disease Control and Prevention (CDC)~\cite{cdc2023}. The app maker installs the app on a large number of mobile devices,
where it collects information on whether a device owner is currently infected, and when, where, and for how long
the device comes in contact with other devices.
Abstractly, one can view the devices as a graph where they are the nodes and their interactions are the edges. Meanwhile, the analyst
wants to use the data to study disease patterns. For instance, it wants to understand the prevalence of superspreaders by evaluating the average number of infected devices in an infected device's neighborhood~\cite{park2020contact,laxminarayan2020epidemiology}. 
\textit{Can we enable the analyst to run such queries and learn their result? Further, can we do it in a way that doesn't require moving device and edge data outside the devices to a centralized location? And further still, can we ensure 
that only the query result is revealed and no new individual device information is learned by any other party?}

This is the problem of \emph{privacy-preserving federated graph analytics}. 
The federated aspect of this problem emphasizes keeping raw data at the devices, in contrast to centralizing 
the data, which is highly 
susceptible to data breaches, especially in bulk~\cite{anthem2015breach,premera2015breach,vojinovic2023data,recent2023breach}.
Meanwhile, the privacy guarantee of the problem emphasizes that an entity should get only the information that it absolutely needs. Thus, an analyst may
learn the query result that is an aggregate across devices. And any device may not learn any more information than it knows locally through its own data and edges.

In this scenario, we consider a threat model where devices and the analyst can be malicious as the data is highly sensitive and compromising devices or the analyst can often happen. Similar to prior work~\cite{roth2021mycelium}, we primarily focus on privacy: we guarantee privacy in the presence of malicious adversaries, but ensure integrity only when all parties are semi-honest. Ensuring malicious integrity would impose much overhead and make the system impractical. Thus we leave ensuring malicious integrity as future work.

As we discuss in related work (\S\ref{s:relwork}), privacy-preserving federated graph analytics is an emerging area of research and displays a trade-off between generality, privacy, and efficiency. 

For instance, Gunther et al.~\cite{gunther2022privacy,gunther2022poster}
have built a system RIPPLE to answer epidemiological questions. However, their system answers aggregation queries only where a device can securely sum its state (e.g., an integer) with its neighbors' state. It doesn't support other secure operations such as multiplication and comparsion. Thus, it cannot answer our motivating query on superspreaders.
RIPPLE also releases  partial sums to the devices, and does not consider the strongest of threat models as a set of its aggregators
are honest-but-curious (but not malicious). 

In contrast, Roth et al.~\cite{roth2021mycelium} have built a
general-purpose system, Mycelium, that assumes a strong threat model where both the devices and a centralized aggregator can be malicious.

However, Mycelium is expensive. For the superspreader query over 1M devices, each device incurs 
8.79~hours of local \cpu time and 5.73~GiB of network transfers.

This paper introduces \sys, an efficient
system for privacy-preserving federated graph analytics in a strong threat model. \sys allows 
an analyst to run simple queries (like the superspreader query) 
that have predicates with a limited set of inputs and outputs, and that evaluate these predicates between a device and its neighbors and then aggregate the results across the devices.
Colo guarantees privacy (only the analyst learns the query result and no entity gets any other intermediate data) while assuming 
malicious devices and a set of $M$ aggregation servers of which a fraction $f$ of them can be malicious. A possible configuration from existing work~\cite{lazar2018karaoke,lazar2019yodel} is to set $M$ to be 40-100 and $f=0.2$. Finally, Colo is scalable and efficient: it supports a large number of devices in the order of a few million, while requiring them to contribute a small amount of \cpu and network.

At a high level, Colo follows a workflow of local evaluation followed by a global aggregation across devices (\S\ref{s:overview}). In the local evaluation, each device evaluates the query  between itself and its neighbors. The global aggregation then aggregates these per-device outputs. 
In this workflow, Colo must address two challenges. First, it must hide node, edge, and topology data during the local evaluation without imposing a large overhead on the devices. 
Second, 
it must aggregate per-node outputs across the population of devices without revealing the intermediate results, and again while being efficient for the devices.

The first challenge of hiding node, edge, and topology data is tricky, especially with malicious devices (\S\ref{s:problem:challenges}).
For instance, say two neighboring devices $v_A$ and $v_B$
want to compute the product $v_A.inf \cdot v_B.inf$ (as needed for the superspreader query), where $inf$ indicates their infection status. Then, a malicious device, say $v_B$, may set its infection status to $v_B.inf = 10^{8}$. 
As a result, the query result secretly encodes $v_A$'s infection status (result is large if $v_A.inf = 1$). 
One may use a general-purpose tool from cryptography to address this issue, but that would be expensive.
Furthermore, even if there were an efficient protocol
for this computation, say that requires  a single interaction between $v_A$ and $v_B$, 
then this protocol must also hide that 
$v_A$ and $v_B$ are communicating, to protect their topology data, i.e., the fact they are neighbors.

Colo addresses the first challenge, particularly, the part of hiding node and edge data,
through a new, tailored secure computation protocol (\S\ref{s:design:localaggregation:1}). 
Colo observes that the query predicates that Colo targets
operate over a limited set of inputs and produce a limited set of outputs. For instance, the legitimate 
inputs and outputs for $v_A.inf \cdot v_B.inf$ are all either zero or one.
Thus, instead of using a general purpose secure computation 
protocol such as Yao's Garbled Circuits~\cite{yao1986generate} that operates over arbitrary inputs and outputs, 
Colo uses a protocol that operates over a limited set of inputs and outputs.
Specifically, one party, say $v_B$, computes all possible legitimate query outputs 
in plaintext, and then allows the other party $v_A$ to 
 pick one of these outputs privately using oblivious transfers (OT)~\cite{rabin2005exchange,chou2015simplest}.
Colo fortifies this protocol against malicious behavior 
of $v_B$
by incorporating random masks, efficient commitments~\cite{brassard1988minimum,grassi2021poseidon}, and 
range proofs~\cite{goldwasser2019knowledge,groth2016size} 
(\S\ref{s:design:localaggregation:1}).

The protocol described above  doesn't yet address the requirement of hiding the topology of devices. 
To hide this data efficiently, that is, the knowledge of who is a neighbor with whom, 
Colo prohibits the devices from directly interacting with each other. 
Rather, Colo arranges for them to communicate via a set of servers, specifically, a set of 40 to 100 servers, where up to 20\% are malicious (\S\ref{s:overview}).
This arrangement enables the servers
to run a particular metadata hiding communication system, Karaoke~\cite{lazar2018karaoke}, that is provably secure and 
low cost for the devices (although with significant overhead for the resourceful servers) (\S\ref{s:design:localaggregation:2}).

Finally, Colo addresses the second challenge of global aggregation across devices through 
straightforward secret sharing techniques while piggybacking on the set of servers (\S\ref{s:overview}, \S\ref{s:design:globalaggregation}).
Specifically, devices add zero sum masks to their local outputs, 
and send shares of these local results to the servers. 
As long as one of the servers is honest, the analyst 
learns only the query output.

We have implemented (\S\ref{s:impl}) and evaluated (\S\ref{s:eval}) 
a prototype of Colo. Our evaluation shows that
for 1M devices connected to at most 50 neighbors each, and for a set of example queries (Figure~\ref{f:query}) which includes the superspreader query, 
a device in \sys incurs less than 8.4 minutes of (single core) \cpu time 
and 4.93~MiB of network transfers.
In contrast, the Mycelium system of Roth et al.~\cite{roth2021mycelium}
requires a device-side cost of 8.79~hours (single core) \cpu time and 5.73~GiB network transfers.
In addition, \sys's 
server-side cost, depending on the query, ranges from 
\$158 to \$1,504 total for \sys's 40 servers  
(the lower number is for the superspreader query).
In contrast, Mycelium's 
server-side cost is
over $\$57,490$ per query. 

Colo's limitations are substantial. In particular, 
it does not handle general purpose queries, rather only those
that evaluate predicates over a bounded set of inputs and outputs.
However, Colo scales to a significant number of devices and is efficient for them, in a strong threat model.
But more importantly, unlike prior work, 
Colo shows that privacy-preserving federated graph analytics can be practical, and 
that the CDC \emph{could} run certain queries over the devices' data while guaranteeing 
privacy in a strong sense, without draining the devices' resources, and without aggressively depleting its own budget (e.g., 
running the superspreader query in a large city every two weeks would cost
around four thousand dollars annually). 
\section{Problem statement}
\label{s:problem}

\subsection{Scenario}
\label{s:problem:scenario}
We consider a scenario consisting of 
a data analyst $\mathcal{A}$ and a large number of mobile devices
$v_i$ for $i\in [0,N)$. For instance, $N = 10^{6}$.

The devices form a graph. 
    As an example, they may run a contact tracing application for COVID-19~\cite{tracetogether,covidsafe,swisscovid} that collects
information on the infection status of the device owners and identities of the devices they come in contact with, that is, their neighbors.
More precisely, a device may have
(i) \emph{node data}, 
    for example, 
    a status variable $inf$ indicating whether the device's owner is currently infected, 
    $tInf$ indicating the time the owner got infected (or null if the owner is not infected), 
    and demographic information such as age and ethnicity;
(ii)
    \emph{edge data}, for example,
        the number of times the device came in contact with a neighbor ($contacts$), the cumulative
        duration ($duration$) of these interactions, and ($location$, $time$, $duration$)
        of each interaction; and,
(iii) \emph{topology data}, for example,
    the list of the device's neighbors.

\begingroup
\setlength{\tabcolsep}{10pt}
\renewcommand{\arraystretch}{1.25}
\begin{figure*}[t]
    \footnotesize
    \centering
    
    \begin{tabular}{
        @{}
        *{1}{>{\raggedright\arraybackslash}b{.08\textwidth}}
        *{1}{>{\raggedright\arraybackslash}b{.80\textwidth}}
        @{}
        }
        Query &  Description\\
        \bottomrule
        Q1 & The total number of infections in an infected participant's neighborhood\\
        \vspace{5pt}
         & \textit{SELECT COUNT(*) FROM neigh(1) WHERE self.inf \& neighbor.inf}\\
         \hline

        Q2 & The amount of time neighbor has spent near infected device if neighbor is infected within 5-15 days of contact with the device\\
          & \textit{SELECT SUM(edge.duration) FROM neigh(1) WHERE self.inf \& neighbor.inf \& (neighbor.tInf $\in$ [edge.lastContact+5 days, edge.lastContact+15 days])}\\
         \hline

        Q3 & The frequency of contact between device and neighbor, if device infected neighbor \\
        & \textit{SELECT SUM(edge.contacts)/COUNT(*) FROM neigh(1) WHERE self.inf \& neighbor.inf \& (neighbor.tInf > self.tInf+2 days)} \\
         \hline

        Q4 & Secondary attack rate of infected devices if they traveled on the subway\\
        & \textit{SELECT SUM(neighbor.inf)/COUNT(*) FROM neigh(1) WHERE self.inf \& onSubway(edge.lastContact.location)}\\
        \hline

        Q5 & The number of secondary infections caused by infected devices in different age groups  \\
        & \textit{SELECT COUNT(*) FROM neigh(1) WHERE self.inf \& neighbor.inf \& (neighbor.tInf > self.tInf+2 days) GROUP BY self.age}\\
        \hline   

        Q6 & The number of secondary infections based on type of exposure (such as family, social, work) \\
        & \textit{SELECT COUNT(*) FROM neigh(1) WHERE self.inf \& neighbor.inf \& (neighbor.tInf > self.tInf+2 days) GROUP BY edge.setting}\\
        \hline

        Q7 & Secondary attack rates in household vs non-household contacts \\
        & \textit{SELECT SUM(neighbor.inf)/COUNT(*) FROM neigh(1) WHERE self.inf GROUP BY isHousehold(edge.lastContact.location)}\\
        \hline

        Q8 & Secondary attack rates within case-contact pairs in the same age group \\
        & \textit{SELECT SUM(neighbor.inf)/COUNT(*) FROM neigh(1) WHERE self.inf \& neighbor.age$\in$[0,100] \& self.age$\in$[neighbor.age-10,neighbor.age+10]}\\

        \bottomrule
    \end{tabular}
    \caption{
    Example graph queries from Mycelium~\cite{roth2021mycelium}
    and the literature on health analytics~\cite{park2020contact,nikolay2019transmission,mossong2008social,laxminarayan2020epidemiology,jing2020household,gudbjartsson2020spread,danon2013social,bi2020epidemiology,adam2020clustering}. We assume that the domain 
    of the inputs to these queries is bounded, for example, $inf \in [0,1]$ and $tinf \in [1,120]$, referring to the days in the latest few months.
    }
    \label{fig:query}
    \label{f:query}
\end{figure*}

\endgroup
The analyst $\mathcal{A}$, say a large hospital or CDC in the context of the contact tracing application, 
wants to analyze the device data by running graph queries.
Figure~\ref{fig:query} shows a few example queries from the literature~\cite{park2020contact,nikolay2019transmission,mossong2008social,laxminarayan2020epidemiology, jing2020household,gudbjartsson2020spread,danon2013social,bi2020epidemiology, adam2020clustering} (these queries are a subset of 
the ones considered in the Mycelium system of Roth et al.~\cite{roth2021mycelium}).
For instance, 
$\mathcal{A}$ may want to learn the number of active infections infected devices have in their neighborhood (Q1). As another example, it may want to learn
how frequently infected devices contact their neighbors who are subsequently infected (Q3). Generally,
these queries perform a local computation at every device and its neighborhood, and then 
aggregate the per-device results. 

\subsection{Threat model}
\label{s:problem:threatmodel}
We assume that the devices
    are malicious, i.e., an adversary can compromise an arbitrary subset of devices.
A compromised device may try to learn information about an 
    honest device beyond what it  knows from its own data.
For instance, if a compromised device is a neighbor of an honest device, then the compromised device already has edge data for their edge (e.g., the location and time of their last meeting); however, the adversary may further want to learn 
if the honest device is infected (neighbor's node data), whether it recently met someone on the subway (neighbor's edge data), and whom it recently came in contact with (neighbor's topology data).

The analyst may also be malicious and want to learn about individual device node, edge, or topology data---information that is more granular than the result of the queries.

Finally, we assume that the adversary may also observe and manipulate network traffic, for instance, in the backbone network, and try to infer
relationships between devices. 

\subsection{Goals and non-goals}
\label{s:problem:goals}

\emparagraph{Target queries.} Ideally, we would support arbitrary graph queries. However, as noted earlier (\S\ref{s:introduction}), 
generality of queries is in tension with privacy and efficiency.
Thus, in this paper we focus on simpler queries such as those in Figure~\ref{f:query} where the aggregations
across devices are SUM, COUNT and AVG operations, and where devices compute simple predicates on a small set of possible inputs 
in their one hop neighborhood (for example, the infection status $inf$ is either zero or one, and 
the time of infection $tinf$ may be in $[1,120]$ referring to the days in the latest few months). 
We note that although these queries are a sub-class of a broad set of queries, they are
important according to the health literature and form a precursor to more sophisticated queries in an analyst's workflow. Besides the limitation on the generality, we don't aim to support interactive queries.

\emparagraph{Privacy (P1).} Private data should always be hidden from the adversary. From the point of view of a device, it may not learn any information beyond its own node, edge, and topology data. In particular, it may not learn any information about a neighbor beyond what is contained in the direct edge to the neighbor.
Similarly, the analyst must only learn the query result.

\emparagraph{Privacy (P2).} Individual devices will cause limited changes to query result. With privacy goal P1, only the query result would be learnt by the adversary. To prevent adversaries from learning (additional) information from the query result, the system must guarantee that individual devices will make bounded contribution to the result.

\emparagraph{Scale and efficiency.} First, our system must support 
a large number of devices, for example, one million.
We assume that these devices have a bounded degree, for example, up to 50 neighbors.
(If the actual device graph has nodes with a larger degree, then they may select a subset uniformly for the query execution.)
Second, in terms of device overhead, we want that to be low, 
for example, minutes of \cpu time and at most a few MiBs in network transfers, which is affordable for mobile devices. 
Meanwhile, if the system employs any servers, e.g., cloud servers, then we want the server-side cost to be affordable. 
In terms of dollar cost of renting the servers, we want the per-server cost to be no more than a few tens of dollars 
(per query over a million devices).

\emparagraph{Correctness (non-goal).} In this paper, we focus on privacy and require the query output to be correct only during periods when all parties (devices and any servers) are semi-honest. Ensuring correctness in the periods of malicious behavior could be very expensive. Thus we leave it as future work.
Nevertheless, the system must ensure privacy at all times.

\begin{figure*}[t]
%\vspace{-3mm}
 {\includegraphics[width=.9\textwidth]{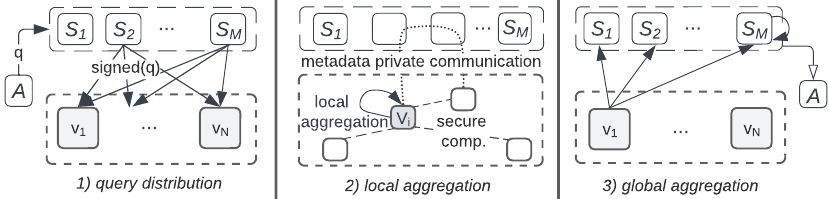}}
\caption{An overview of \Sys's query distribution, local aggregation, and global aggregation phases of query execution. The dotted line in local aggregation depicts metadata-hiding communication, and the dashed line depicts secure computation.
}
\label{f:overview}
\label{fig:overview}
\label{f:arch}
\label{fig:arch}
\end{figure*}

\subsection{Challenge and straw man solutions}
\label{s:problem:challenges}
Meeting these goals, particularly, privacy and efficiency, when the parties can behave maliciously is hard. We explain this point by discussing possible approaches and attacks below. We will use  query Q1 (Figure~\ref{f:query}), calculating the total number of infected direct neighbors of all infected devices, for illustration purposes.

\emparagraph{Centralized server.} One approach is to assume a trusted central server and ask devices to upload their data in plaintext. For example, each device uploads the node data ($inf$) along with its neighbor list, and edge data for all the edges. The server can  evaluate the query using a graph analytics system such as GraphX~\cite{gonzalez2014graphx} as the data is in a single location in plaintext.

This approach, however, breaks the privacy goal P1 immediately. The server sees individual device data and learns more than the query result.

This approach also breaks the privacy goal P2. Though the devices do not learn more than they should in this approach, malicious devices can execute subtle attacks to cause a victim device contribute more than desired to the query result and thus leak information. For the example query Q1, the server needs to compute $\sum_{v_A,v_B} v_A.inf\cdot v_B.inf$ for all pairs of neighbors $(v_A, v_B)$ to calculate the total number of infections in infected devices' neighborhood. A malicious device $v_0$ can send a large constant, e.g., $v_0.inf=10^6$ to the server. As a result, if an honest device $v_1$ is $v_0$'s neighbor, the product $v_1.inf\cdot v_0.inf$ will contribute a large value in the aggregation and the adversary can infer whether $v_1.inf$ is 0 or 1 by inspecting the query result. As another example of breaking P2, a set of malicious devices can claim that a victim device is their neighbor. Again for the example query, if $10^6$ malicious devices claim $v_1$ is their neighbor and set their $inf$ to be 1, the server will compute $\sum_{v_A,v_B} v_A.inf\cdot v_B.inf + \sum_{10^6} 1\cdot v_1.inf$. This computation again amplifies the victim device $v_1$'s data in the aggregate output.

To hide the data from the server (privacy goal P1), one approach is to use tools from cryptography such as homomorphic encryption~\cite{rivest1978data,fan2012somewhat} or secret sharing~\cite{blakley1979safeguarding, shamir1979share}. Meanwhile, to meet privacy goal P2, the server will have to run logic to verify devices' (secret) inputs to prevent malicious devices from uploading arbitrary data.
However, this approach doesn't scale to a large number of devices. For example, for Q1,
the server will need to perform at least $O(N)$ multiplications ($N$ is the number of devices), and also verifications of devices' data. Indeed, as we discuss in related work (\S\ref{s:relwork}), we are not aware of any prior work that follows the centralized server approach when the threat model assumes malicious parties.

\emparagraph{Federated analytics.} Another  approach to the problem is the idea of keeping the data federated. 
Each device keeps its raw data local and first communicates with its neighbors to perform local computation (e.g., aggregation at the neighborhood level), before uploading
    this local result to a server for aggregation. 
The federated approach has distinct advantages relative to the centralized approach:
(i) it allows devices to feel more trust in the system as they keep possession of their data,
(ii) it doesn't risk bulk data breaches as devices do not collectively centralize their data, 
and (iii) 
    it handles data updates naturally as each device computes locally and can use its latest data.
However, efficiency is still challenging.

The state-of-the-art  system that follows the federated approach, 
Mycelium~\cite{roth2021mycelium}, assumes malicious devices and a single untrusted server and scales to a large number of devices. 
However, even for our simple example query Q1 and for 1M devices, a Mycelium device spends 8.79 hours in \cpu time and 5.73~GiB in 
network transfers (\S\ref{s:eval}). The server is also expensive: it spends 1304~h \cpu time and 5737~TiB network transfers.
One core challenge is that during the local computation (where a device performs local aggregation over its own and its neighbors' data), 
    Mycelium still needs to protect against the subtle attacks against privacy goal P2 mentioned above (e.g., one malicious neighbor inputting an arbitrary value into the aggregation)---attacks for which Mycelium uses general-purpose (and expensive) homomorphic encryption~\cite{yagisawa2015fully} and zero-knowledge proofs~\cite{groth2016size}.
Besides, Mycelium must protect against arbitrary behavior of the untrusted server, which may  try to 
manipulate the aggregation, e.g., add a device's data a large number of times to learn it.
\section{Overview of Colo}
\label{s:arch}
\label{s:overview}

Given
the unique advantages of the federated approach relative to the centralized approach, 
Colo follows the former. To facilitate federation, Colo
relies on a set of servers, for example,
40-100 servers, $S_i$ for $i\in [0,M)$ (Figure~\ref{f:arch}). 
These servers may be in separate administrative domains (e.g., 
Azure versus Amazon AWS versus Google Cloud Platform, and different geographical zones 
within these cloud providers).
Colo assumes that at most
$f = 20\%$ of these servers may be compromised by the adversary. That is, an adversary may compromise, e.g.,
up to 8 servers when the number of servers is 40. $M$ and $f$ are both configurable and we followed prior works~\cite{lazar2018karaoke,lazar2019yodel} to use this default setting.

The devices connect to the servers in a star topology, 
where the central hub of the star is the 
set of servers and a spoke connects to a device. Thus, 
the devices do not  directly interact with each other; rather, they communicate
 via the servers. 
This design is necessary for two reasons. First, the devices typically do not have each other's IP address. Second, 
the communication via the servers helps hide the devices' topology data.

Colo has a one-time \emph{setup} phase, and 
three phases of
    \emph{query distribution}, 
    \emph{local aggregation}, and 
    \emph{global aggregation} for query execution (Figure~\ref{f:overview}).
    
\begin{figure*}[t]
\hrule
\medskip
{

\vspace{-0.02ex}
\textit{Setup (one time)}
\vspace{-0.5ex}
\begin{myitemize}
    \item The servers $S_i$, $i \in [0, M)$, generate 
    a set of keys for a zero-knowledge proof (ZKP) scheme using a MPC protocol~\cite{zokrates2023zokrates}.
    
    \item The servers distribute the keys to the devices. A device downloads the set of keys from one server and hashes of these keys from the others to defend against attacks from malicious servers (a malicious server can distribute malicious keys). The devices  stores the keys locally.
    
\end{myitemize}

\vspace{-0.04ex}
\textit{Query distribution}
\vspace{-0.5ex}
\begin{myenumerate2}
    \item The analyst $\mathcal{A}$ submits a query $q$ to all the servers.
    \item The servers verify the query and sign it. Each server broadcasts its signed query to all devices to start 
    query execution.
\end{myenumerate2}

\vspace{-0.4ex}
\textit{Local aggregation}
\vspace{-0.5ex}
\begin{myenumerate2}
\setcounter{enumi}{2}

    \item \label{l:localaggregationtopology} The servers run the Karaoke system~\cite{lazar2018karaoke} to enable the devices to communicate anonymously with each other.
    
    \item \label{l:localaggregation}  Each device initializes its local result of query execution to 0 and 
    runs the local aggregation secure computation protocol (\S\ref{s:design:localaggregation:1}; Figure~\ref{f:localaggregation}) 
    with its neighbors over the Karaoke system~\cite{lazar2018karaoke} to obtain a share of its local result.
    
\end{myenumerate2}

\vspace{-0.4ex}
\textit{Global aggregation}
\vspace{-0.5ex}

\begin{myenumerate2}
\setcounter{enumi}{4}
    \item Each device secret shares its own share of its local result with the servers.
    
    \item Each server sums all secret shares it receives. Each server sends its sum to the analyst and the analyst who adds the sum across the servers to reconstruct the global result.
\end{myenumerate2}
}
\hrule
\caption{
A high-level description of Colo's phases.}
\label{f:system_design}
\end{figure*}

In the setup phase, the servers generate and distribute keys for a cryptographic protocol used in the local aggregation phase. Specifically, 
they generate the proving and verification keys for a zero-knowledge proof (ZKP) scheme~\cite{zokrates2023zokrates}.

In the query distribution phase (the leftmost diagram in Figure~\ref{f:overview}), the analyst $\mathcal{A}$ specifies the query, say $q$, and sends it to each of the servers. 
Since $\mathcal{A}$ can be malicious (\S\ref{s:problem:threatmodel}) and may write a query that tries to infer a single device's data, the servers validate
$\mathcal{A}$'s query. For example, if $q$ contains a WHERE clause of the form \textit{WHERE self.ID = xxx}, then the servers reject it. In general, checking whether a query leaks information about an individual device is an open and difficult research problem~\cite{gehrke2011towards,li2023private,day2016publishing,ding2018privacy}.
Colo does not aim to solve it and assumes that the servers have a list of certified queries that are allowed. Once each server validates the query, it signs and broadcasts it to the devices. A device starts the next phase 
after validating enough signatures, that is, more \changebars{than}{that} the threshold of servers that can be compromised.

In the local aggregation phase (the center diagram in Figure~\ref{f:overview}), each device evaluates the query in its neighborhood. For instance, for the query Q1, \textit{SELECT COUNT(*) FROM neigh(1) WHERE self.inf \& neighbor.inf}, each node computes
the local count of infected neighbors in its neighborhood. More precisely, each node computes the count of infected neighbor (either zero or one) for each edge, and then adds the results across the edges. Recall that our goal is to ensure that a device learns no more information than it needs to (\S\ref{s:problem:goals}). Thus, a device uses a secure computation protocol with its neighbor such that at the end of the protocol the two parties receive \emph{secret shares} of the result of the computation.
This protocol admits malicious devices; for instance, a malicious neighbor is prevented from supplying an arbitrary input such as setting its $inf = 10^{6}$.

Secure computation hides node and edge data; however, an adversary that observes network traffic can infer topology  by monitoring who is performing secure computation with whom. Thus, the devices in the local aggregation phase execute secure computation over a metadata-hiding communication network, 
particularly, the Karaoke system~\cite{lazar2018karaoke}. The servers facilitate and run this system.

Finally, in the global aggregation phase (the right diagram in Figure~\ref{f:overview}), the devices send their results from the local aggregation to the servers, who aggregate them. Specifically, each device secret shares its result with the servers, who locally add the shares they receive from the devices. The servers send the result of their local computation to the analyst who combines these outputs across the servers to obtain the query result.

Colo's architecture (with the star topology where the hub is a set of servers) offers advantages for the privacy versus efficiency tradeoff of Colo.
First, the servers form a natural infrastructure to instantiate a metadata-hiding communication system such as Karaoke~\cite{lazar2018karaoke,lazar2019yodel}.
Second, this architecture keeps the cost of global aggregation low by enabling servers to perform only local computations in the global aggregation phase. In contrast, if we assumed a completely untrusted server as the hub (as in Mycelium~\cite{roth2021mycelium}), then this untrusted server would have to use a sophisticated verification protocol to 
prove that it performed the global aggregation correctly (for instance,
prove that it did not add a victim device's input many times).
There would also be no resourceful infrastructure to run a metadata-hiding communication system.

Nevertheless, there are still several challenges in instantiating this architecture, one of which is the efficiency of the secure computation piece in the local aggregation phase. This protocol has to be particularly efficient in terms of network transfers 
because this overhead gets exacerbated when 
secure computation messages are routed through the metadata-hiding network. 
We next go over the design details of Colo and how it addresses the various challenges.
\section{Design details}
\label{s:design}
This section describes the details of Colo. Figure~\ref{f:system_design} provides a high-level
overview and how the phases connect with each other.

\subsection{Setup (key generation)}
\label{s:design:setup}

The first challenge Colo must deal with is the overhead of key generation and distribution.
As we will describe later (\S\ref{s:design:localaggregation:1}), devices in Colo's local aggregation phase 
need to generate an array $T$ of values and prove that each value is within a range $[L, U]$. 
The challenge is that the efficient and popular zero-knowledge proof (ZKP) schemes such as that of Groth~\cite{groth2016size} bind 
the proving and verification keys to the statement (the circuit) the prover is proving. 
That is, 
this circuit is a function 
of the length $len(T)$ 
of array $T$ in our case. Unfortunately, the $len(T)$ further depends on the analyst's query $q$, and may be different for different queries.

On the one extreme, Colo could use a single proving and verification key that works for $len(T) = 1$. Then, when proving the properties
of an array $T$ with $len(T) > 1$, it could break down the array into singletons and generate a separate proof for each.
The benefit is  efficiency of key generation and distribution as Colo's servers will need to 
generate and distribute one key pair. However, the penalty is during query execution as Colo's devices would have to generate many proofs. 
For instance, for a query with $len(T) = 1024$, the \cpu time doubles and
the size of the proof increases from 192~B to 192~KiB, relative to generating a single, holistic proof. 
This network overhead is significant because each message 
goes through the metadata-hiding communication network (\S\ref{s:overview}).

On the other extreme, Colo could generate, say, 1024 keys for all possible lengths of $T$ between 1 and 1024.
In this extreme, key generation and distribution is expensive; for instance, the set of keys which has to be shipped to each device 
(and stored there) will total 52.4~GiB.
However, the proof generation is optimal with a single proof for the array $T$, thereby lowering 
\cpu and network overhead during query execution.

Colo resolves this tension between overhead of key and proof generation by generating a key set
$\{key_{2^0}, key_{2^1}, \ldots, key_{2^{\log(len(T))-1}}\}$ containing $\log(len(T))$ keys for all powers of two between $1$ and anticipated maximum length $len(T)$, for example, 1024.
Now, the size of the key set is manageable, e.g., 102~MiB, and during query execution devices generate no more than $\log(len(T))$ proofs.

\subsection{Query distribution}
\label{s:design:querydistribution}
As mentioned in the overview (\S\ref{s:overview}), the analyst $\mathcal{A}$ specifies 
a query $q$ and gives it to the servers, who validate it. Here, we elaborate on how $\mathcal{A}$ specifies the query.

Each query has two components: (i) a SQL query similar to the examples we have discussed (Figure~\ref{f:query}), and (ii) a transformation function, \emph{PreProcess}, that transforms the raw data at a device into a form needed by the SQL query. One can view the \emph{PreProcess} function as implementing preprocessing or cleaning of data and the SQL query as the analysis over the prepossessed data.

As noted earlier (\S\ref{s:problem:goals}), Colo does not support arbitrary SQL queries. 
Rather, it targets simple queries of the form
\textit{SELECT AGG-OP(g(self.data, neighbor.data)) FROM neigh(1) WHERE h(self.data, neighbor.data)}, where
AGG-OP is the SUM, COUNT, or AVG aggregation operation, $g$ is a predicate on 
the data of two neighbors (including their node and edge data), and $h$
is a filter that runs over the same data 
to compute whether a particular edge will participate in aggregation or not.
If $\mathcal{A}$ wants to run a query with a 
GROUP BY operation, $\mathcal{A}$ transforms it 
into several sub-queries and submits them to the servers separately. 
For example, $\mathcal{A}$ converts Q5 in Figure~\ref{f:query} into a series of queries such as \textit{SELECT COUNT(*) FROM neigh(1) WHERE 
20 < self.age < 30} for the different age groups.

The \emph{PreProcess} function specifies how a device
should translate its raw data into the attributes accessed by the SQL query. 
For example, the analyst $\mathcal{A}$ may want to execute Q3 in Figure~\ref{f:query} (\textit{SELECT SUM(edge.contacts)/COUNT(*) FROM neigh(1) WHERE self.inf \& neighbor.inf \& (neighbor.tInf > self.tInf+2 days)}) 
for all infectees during a recent time period, e.g., March 2023, as 
suggested in the literature~\cite{gudbjartsson2020spread}. 
The PreProcess function should specify how to derive the following three attributes for the SQL query: i) device infection status $inf$ as a binary number; ii) the infection timestamp $tInf$ as an integer from 1 to 30 to represent March 1 to 30; iii) $edge.contacts$, which is the number of interactions two neighbors have had, as a bounded integer, for example,
80 as in the epidemiology literature~\cite{laxminarayan2020epidemiology}.
These constraints (bounds) are necessary as otherwise malicious devices can supply arbitrary inputs. 

Once the servers receive $\mathcal{A}$'s query,
they verify it by matching it to a list of certified queries. The servers then sign and broadcast the query; if a device verifies 
signatures more than the fraction of servers that can be compromised, it starts query execution.

\subsection{Local aggregation}

\subsubsection{Hiding node and edge data}
\label{s:design:localaggregation:1}
\begin{figure*}[t]
\hrule
\medskip
{
%\fontsize{9.5}{10.8}\selectfont
\textbf{Local aggregation protocol of \sys}

\vspace{-0.02ex}
%\textit{Commit step}
\vspace{-0.5ex}
\begin{myenumerate2}

    \item Each device receives a query $q$ containing a function $PreProcess$, a local computation $F$, and a bound $Bound$. $PreProcess$ contains information on how to convert a device's raw data into inputs for $F$. It also specifies
    the set of valid values for each input parameter to $F$.
    The bound $Bound$ is the range of valid outputs of $F$.
    
    \item Each device participates in this local aggregation protocol. Denote $A$ as the device and $B$ as a neighbor.
    
    \item Each neighbor $B$ of $A$ does the following locally:
    \begin{myenumerate}
    
        \item (Generate mask) Samples an element $r$ in a field $\mathbb{F}$ uniformly randomly. 
        
        \item (Enumerate all inputs of $A$) Generates an array $s$ containing all possible inputs of $A$ based on
        $PreProcess$. 
        \item (Compute outputs) For every $s[i]$, computes $T[i] = F(s[i], B_{in})$, where $B_{in}$ is $B$'s input.
        \item (Mask outputs and commit to them) For every $T[i]$, computes $T'[i] = T[i] + r$. It then samples $R[i]$ and generates commitment $CM[i] = Commit(T'[i], R[i])$.
        \item (Prove bound of outputs) Generates a ZKP that it knows the opening of all commitments and a mask such that each committed value is the addition of the mask and some value bounded by $Bound$.
       
    \end{myenumerate}

    \item Each neighbor $B$ sends commitments $CM$ and the ZKP to $A$ which verifies the ZKP.
    \label{f:localaggregation:cmzkp}

    \item (Function evaluation) $A$ runs OT with each neighbor $B$ to retrieve $(T'[j], R[j])$, where $s[j]$ is $A$'s input to $F$. The device $A$ verifies the opening to the commitment, that is, it verifies $Commit(T'[j], R[j])$ equals $CM[j]$ received in the previous step.
    \label{f:localaggregation:ot}

    \item If the verifications above pass, A adds $T'[j]$ to its local aggregation result. Its neighbor device $B$ adds $-r$ to its local aggregation result.
    
\end{myenumerate2}
}
\hrule
\caption{%
\Sys's local aggregation.}
\label{f:localaggregation}
\end{figure*}

Recall from the overview (\S\ref{s:overview}) 
that the goal of local aggregation is to enable a device to compute the query locally 
just on its neighborhood. This computation further breaks into evaluating the query for
every neighbor edge of a node. 
That is, a node needs to evaluate 
a function $F = g \circ h (self.data, neighbor.data)$ with each of its 
neighbors and compute $\sum_{neighbor} F(self.data, neighbor.data)$. 
As an example, for query Q1 the function $F$ equals
$F(self.data, neighbor.data)= self.inf \cdot neighbor.inf$.

For the moment, assume that we do not need to hide the topology data at the devices (we will relax this assumption in 
the next subsection).
Then, a natural option for computing $F$ is to use a two-party secure computation protocol such as Yao's garbled circuit (Yao's GC)~\cite{yao1982protocols}. 
The neighbor, say, $v_B$, could act as the garbler, generate a garbled circuit, send it to the other node, say, $v_A$, 
who would act as the evaluator to obtain the result. 
Since the two nodes must not obtain the result of $F$ in plaintext, the two nodes may compute
$y = F(r_{v_A}, r_{v_B}, v_A.inf, v_B.inf) = 
(v_A.inf \cdot v_B.inf) + r_{v_A} + r_{v_B}$,
where $r_{v_A}, r_{v_B} \in \mathbb{F}$ are uniformly sampled masks supplied by the two parties to hide the output from each other. (The field $\mathbb{F}$ could be $2^{64}$, for example.) 
At the end of the protocol,
$v_B$ may store $-r_{v_B}$ as its output, while $v_A$
may store $- r_{v_A} + y$ as its output.

The challenge is in preventing malicious behavior efficiently. To protect against a malicious neighbor (garbler $v_B$), 
Colo would have to use a version of Yao's GC that employs techniques such as cut-and-choose~\cite{lindell2007efficient,mohassel2013garbled,lindell2012secure} 
that prevent
a garbler from creating arbitrary circuits, for example, an $F'$ that computes,
$(v_A.inf \cdot 10^{6}) + r_{v_A} + r_{v_B}$. These general-purpose primitives are expensive because they are not tailored for the queries and they offer more than we need: malicious integrity is not our goal ($\S\ref{s:problem:goals}$).

Observe that the predicates $F$ that appear in Colo's target queries have bounded inputs and outputs.
For example, Q1 has two possible inputs of 0 and 1 for $v_A.inf$ and thus two
possible outputs of $0$ and $1$. As another example, if we take Q3 in Figure~\ref{fig:query} and assume $tInf$ has 30 possibilities (for 30 days), 
then the number of possible inputs for $v_A$ is sixty. 
That is, $v_A$'s
input pair $(inf, tInf)$ can range from the case $(0,0)$ to the case $(1,29)$.
Corresponding to each of these inputs, an output $edge.contacts$ may be a value in the range $[0,79]$.

Leveraging this observation, the neighbor in Colo ($v_B$ above) \emph{precomputes} 
outputs for all possible inputs of $v_A$ into an
array $T$ instead of generating them at \emph{runtime} inside Yao's GC. 
One can view this arrangement as making $v_B$ 
commit to the outputs without looking at $v_A's$ input. For instance, 
for Q3, $v_B$ would generate a $T$ of length 60 where each entry is 
in the range $[0,79]$. Then, $v_B$ can arrange for $v_A$ to get one of these values obliviously.

\emparagraph{Protocol details.}
Figure~\ref{f:localaggregation} shows the details of Colo's protocol. 
For a moment, assume that the nodes $v_A$ and $v_B$ are honest-but-curious. Then, for every possible input of $v_A$, the neighbor $v_B$ generates one entry of array $T$. It also adds a mask, $r \in \mathbb{F}$, to each entry of $T$.
That is, after generating the array $T$, $v_B$ offsets each entry by the same mask and computes $T'[i]=T[i] + r, i\in [0, len(T))$, where $r$ is private to $v_B$.
The node $v_A$ then obtains one of the entries of the table corresponding to its input using 
1-out-of-$len(T)$ oblivious transfer~\cite{chou2015simplest,rabin2005exchange}.
Finally, $v_A$ uploads $T[j]+r$ for the global aggregation, and $v_B$ uploads $-r$ to cancel the mask.

To account for a malicious neighbor $v_B$ who tries to break our privacy goal P2 ($\S\ref{s:problem:goals}$), Colo's protocol 
adds a zero-knowledge range proof~\cite{groth2016size}. Specifically, $v_B$
commits to the values in $T$ and proves that each value is bounded and that each value is offset by the same private 
mask $r$. Recall from the setup step (\S\ref{s:design:setup}) that the keys for the ZKP are specific to length $len(T)$
of array $T$. Thus, $v_B$ splits up proof generation as needed depending on the binary representation of $len(T)$.
After generating the proofs, $v_B$ sends them to $v_A$ along with one entry $T'[j]$ of $T'$ (using OT)
as well as the opening for the commitment to $T'[j]$. Now,
instead of using an OT protocol secure only against an honest-but-curious sender, Colo's protocol 
switches to an OT that defends against a malicious sender ($v_B$)~\cite{chou2015simplest,orru2016actively}.

Colo's protocol is not general purpose and is not meant to replace Yao's GC. In particular, its overhead is proportional to the 
input domain size of the predicate $F$. Thus, it only works for small input domains. 
However, its performance benefits are very significant, and allows the devices to keep their overhead low.
First, node $v_B$ evaluates the predicate $F$ in plaintext rather than inside Yao's GC.
Second, the protocol uses ZKP for range proofs, a computation which has received significant attention in the literature~\cite{bunz2018bulletproofs,chung2022bulletproofs}.
Third, and similar to the point above, the OT primitive is also a basic secure computation primitive that 
has received much attention on performance optimizations~\cite{chou2015simplest, orru2016actively,mansy2019endemic,roy2022softspokenOT}.

\subsubsection{Hiding topology}
\label{s:design:localaggregation:2}
The local aggregation 
protocol so far hides node and edge data; however, an adversary that can monitor network 
traffic can infer who is running secure computation with whom and 
thus figure out the topology data for the devices. 
As noted earlier (\S\ref{s:overview}), 
Colo protects this information by enabling the devices to communicate over a metadata-hiding communication system.
In particular, Colo uses a state-of-the-art system called Karaoke~\cite{lazar2018karaoke}. Here, we briefly review Karaoke as it has significant 
implications on Colo's server-side overhead.

At a \changebars{high level}{high-level}, Karaoke relies on the concept of \emph{dead drops}---a pseudorandom location (mailbox) at a server.
Say device $v_{A}$ wants to send a message to $v_{B}$, and they have a shared secret $k_{AB}$ to facilitate this communication (e.g., they can agree on this secret 
when they become neighbors). To send a message $msg$ to $v_B$, the device $v_{A}$ 
first derives a dead drop and picks a server $S_{drop}$ hosting the dead drop, using $k_{AB}$ as the seed of a pseudorandom generator~\cite{yao1982theory,blum2019generate,bernstein2008chacha}. 
The device $v_{A}$ then drops (writes) the message at the dead drop, while device $v_{B}$ retrieves the message from the dead drop.

Of course, the idea of dead drops alone is insufficient, as the dead drop server $S_{drop}$
can see which devices are communicating with the same dead drop. To fix this issue, Karaoke makes the devices access the dead drops via servers, similar to a parallel
mixnet~\cite{chaum1981untraceable}. 
 Suppose the key pairs of two devices $v_A, v_B$ are $(pk_A, sk_A)$, $(pk_B, sk_B)$, and the key pairs for the 
 $M$ servers $S_i, i\in [0, M)$'s are $(pk_i, sk_i)$. Then, when the device $v_A$ wants to send a message to a dead drop server, it chooses a uniformly random path of $m$ servers, say, $S_0, \ldots, S_{m-1}$, to route the message. Specifically, $v_A$  onion encrypts~\cite{chaum1981untraceable} the message as
$Enc_{pk_{0}}(\ldots Enc_{pk_{m-1}}(Enc_{pk_{drop}}(msg)))$ and sends it to the first hop on the path.
When a server on the path receives a message, it peels one layer of the onion and forwards the remaining message to the next server, until the drop 
server receives it. To protect against malicious servers dropping messages to compromise privacy, the servers add dummy traffic (noise messages to random dead drops).

Karaoke adds both significant \cpu and network overhead to the servers ($\S\ref{s:eval}$). 

Despite this overhead, we pick Karaoke for several reasons. 
First, the overhead is at the server-side, where there is more tolerance for  overhead relative to the devices.
Second, Karaoke can scale to millions of devices as needed for our scenario.
Third, it defends against malicious servers who may duplicate or drop traffic to learn access patterns at the dead drops. 
And, fourth, it provides a rigorous guarantee of
$(\epsilon, \delta)$ differential privacy. 
 For instance, it makes the case that $v_A$ and $v_B$ are communicating appear indistinguishable from the case when they are idle. 
 In particular,
with a route length $m = 14$, assuming 80\% of the servers
are honest, it guarantees indistinguishability 
with a probability of 
$1-2 \cdot 10^{-8}$.

\subsection{Global aggregation}
\label{s:design:globalaggregation}

The global aggregation phase is the last phase in query execution. Recall that
at the end \changebars{of }{}the local aggregation phase, each device has 
a local result, say $y_i$. For instance, $y_i$ equals
the output of secure computation $T'[j]$ in Figure~\ref{f:localaggregation} or the mask $r$ added 
by a device that constructed the array $T'$. 
The goal of the global aggregation phase is to aggregate these outputs across devices to wrap up the execution of the query.
For this aggregation, each device secret shares (using additive shares in the field $\mathbb{F}$) its output $y_i$ and sends one share to each server. 
Each server then locally computes $\sum_i [y_i]$, where $[y_i]$ is a share it receives. 
The analyst finally aggregates the results across the servers, that is, computes $\sum_{edge}(T[j] + r) + \sum_{edge}(-r)$, \changebars{canceling}{cancelling} out the random masks, and obtaining the query output.

This simple global aggregation protocol ensures that the local 
aggregation results of an honest device get aggregated with 
the results of other honest devices exactly once. 
For instance, if a malicious server drops or duplicates a share $[y_i]$
supplied by an honest device, then the analyst $\mathcal{A}$ learns a uniformly random output, because the shares of $[y_i]$
at an honest server will not cancel out.
Naturally, this protocol guarantees privacy as long as one of the servers is honest, which 
is satisfied by our assumption of only up to $f=20\%$ of servers being 
malicious (and the rest are honest).
\section{Implementation}
\label{s:impl}

We have 
implemented a prototype of Colo in C{++}.
Our prototype is approximately 2,000 lines of code on top of existing frameworks and libraries. 
In particular, we use CAF~\cite{chs-rapc-16}, an event-driven framework used commonly for building distributed systems. We instantiate Colo's servers and devices as actors in this framework. The source code will be made available on Github.

\emparagraph{Server-side details.}
We use Groth16~\cite{groth2016size} as our underlying zkSNARK (ZKP) scheme and we use BLS12-381 as its underlying curve 
for a 128-bit security. 
In the setup phase, our prototype generates a set of ZKP keys in MPC using ZoKrates~\cite{zokrates2023zokrates}. 
This protocol ensures that the randomness used to generate the keys is not revealed to an adversary, and that the generated keys are correct.
In the local aggregation phase, servers run Karaoke~\cite{lazar2018karaoke} to provide metadata-hiding messaging. We follow Karaoke's default configuration to generate noise messages to achieve $\epsilon=\ln 4,\delta=10^{-4}$ differential privacy after 245 rounds of message losses.
In the global aggregation, servers accept devices' additive secret shares of local results and reveal the summation of these results to the analyst. Our prototype uses $2^{64}$ as the underlying field $\mathbb{F}$ for secret shares.

\emparagraph{Device-side details.}
In the query distribution phase, the devices take as input two 
functions in C{++}, one for the predicate $F$ and the other for the \emph{PreProcess} transformation function (\S\ref{s:design:querydistribution}).
The devices also take as input a map object
indicating the range of each 
attribute accessed by the query. 
For the local aggregation phase, recall (Figure~\ref{f:localaggregation}) that devices need
cryptographic tools, particularly, a commitment scheme, a ZKP scheme, and an OT protocol.
For commitments, our prototype uses a
ZKP-friendly scheme called Poseidon~\cite{grassi2021poseidon}, by
importing the ZKP circuit for this commitment from neptune~\cite{neptune} into
ark\_groth16~\cite{arkgroth16}.
It also implements a range proof to prove that the values in the array $T$ are in the range bound. 
For OT, we use the simplestOT protocol~\cite{chou2015simplest} implemented in libOTe~\cite{libOTe}.
For enabling the devices to participate in Karaoke (metadata-hiding communication), we use the onion encryption scheme implemented in libsodium~\cite{libsodium}.
Finally, for global aggregation, the devices simply sample random numbers in $2^{64}$
to generate the additive secret shares of their local results.

\section{Evaluation}
\label{s:eval}
\begin{figure}[t]
    \footnotesize
    \centering
    
    \begin{tabular}{
        @{}
        *{1}{>{\raggedright\arraybackslash}b{.12\textwidth}}%
        *{1}{>{\raggedleft\arraybackslash}b{.12\textwidth}}  % 
        *{1}{>{\raggedleft\arraybackslash}b{.11\textwidth}}  % 
        @{}
        }
        Application & Dataset &  \# of nodes \\
        \bottomrule
        Social & ego-Facebook & 4,039 \\
        Communication & Enron email & 36,692 \\
        Collaboration & DBLP & 317,080 \\
        Collaboration & live journal & 3, 997,962 \\
        \bottomrule
    \end{tabular}
    %\normalfont\selectfont
    \caption{Commonly used graph datasets from 
    the Stanford Large Network Dataset Collection (SNAP)~\cite{snapnets}.}
    \label{fig:dataset}
    \label{f:dataset}
\end{figure}

\begin{figure*}[t]
\centering
%\hspace*{-2.5em}
\begin{subfigure}[t]{0.4\textwidth}{
\centering
\includegraphics[width=2.5in]{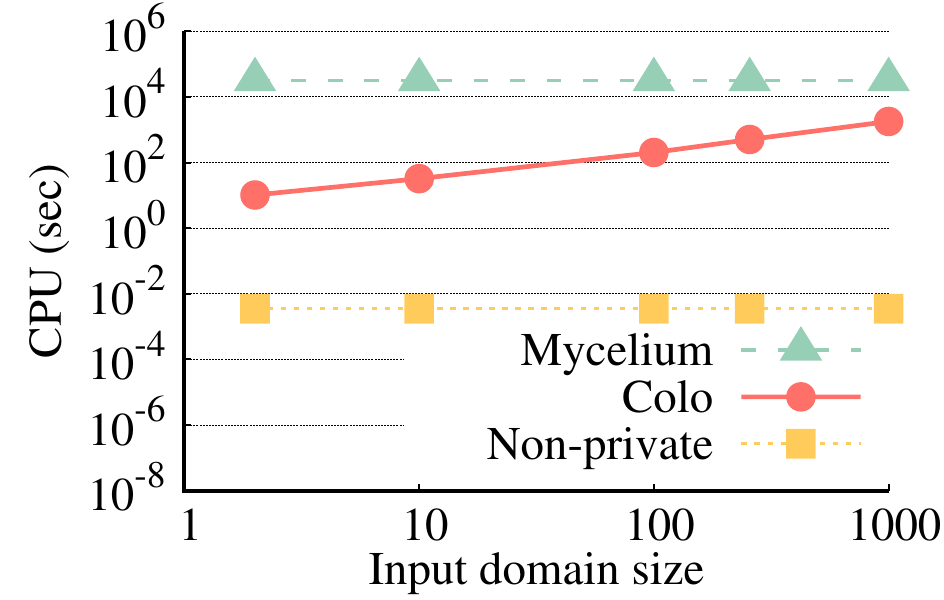}
\caption{Device \cpu}
\label{f:client_cpu_vs_tt}
}
\end{subfigure} %\hspace*{\fill} 
\begin{subfigure}[t]{.4\textwidth}{
\centering
\includegraphics[width=2.5in]{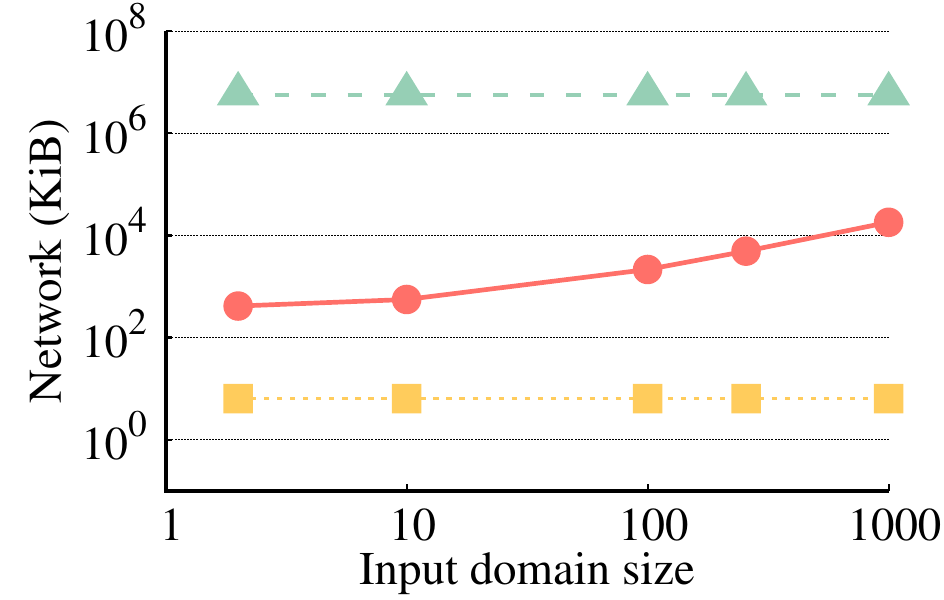}
\caption{Device network}
\label{f:client_net_vs_tt}
} %\hspace*{\fill}
\end{subfigure}
\caption{
Device-side cost with a varying input domain size $len(T)$ of the query predicate. Queries 
Q1, Q2, Q4, Q7 from Figure~\ref{f:query} have 
$len(T)=2$, Q3, Q5, Q6 have $len(T)\in [60,240]$, and Q8 has $len(T) = 240$.} 
\label{f:client_vs_tt}
\label{fig:client_vs_tt}
\end{figure*}

Our evaluation focuses on highlighting the device- and server-side overhead of Colo, for a variety of queries, and for different underlying graphs (the number of nodes and edges, and the degree of these nodes).
A summary of our main results is as follows:

\begin{myitemize}

\item For 1M devices connected to up to 50 neighbors each, and for all example queries (Figure~\ref{f:query}), 
\sys's per-device cost is less than 8.4 min of (single core) \cpu time 
and 4.93~MiB of network transfers.

\item For the same number of devices and their neighbors as above, \sys's 
server-side cost is $\$3.95$ to $\$37.6$ per server 
(or \$158 to \$1,504 total for the 40 servers), depending on the query. 

\item \Sys's overheads for the example queries are much lower than the state-of-the-art Mycelium, whose device-side cost is 8.79~hours (single core) \cpu time and 5.73~GiB network transfers, per query, and the server-side cost is
$\$57,490$ per query. Thus, \sys brings privacy-preserving federated graph analytics into the realm of affordability for certain types of queries.

\end{myitemize}

\emparagraph{Baselines.}
We compare Colo to two baseline systems: a federated non-private baseline and the state-of-the-art  Mycelium~\cite{roth2021mycelium} for privacy-preserving federated graph analytics. 

Our non-private baseline has a coordinator server. It forwards messages between devices as they typically do not have public IP and cannot interact directly. Using the coordinator, devices run graph queries in a federated manner (but they are not privacy-preserving): each neighbor $v_B$ of $v_A$ sends its data to $v_A$ in plaintext, who then performs local computation and uploads the result to the coordinator. The coordinator aggregates the results across devices. This baseline is non-private as devices see their neighbors' data, and the coordinator sees 
the local results computed at the devices. Besides, this baseline does not consider malicious devices.

For a privacy-preserving baseline, we use the state-of-the-art Mycelium~\cite{roth2021mycelium}.
However, this is challenging as Mycelium only has a partial public implementation~\cite{myceliumgithub}.
In particular, the device-side zero-knowledge proof
implementation does not include the full proof, and the server-side code does not include the aggregation of computation across devices.
To work around these constraints, we report lower-bound estimates for Mycelium. Briefly, however,
for device-side \cpu, we use a lower-bound estimate of ZKP since it is the dominant device-side cost (Mycelium performs homomorphic operations over ciphertexts and proves that
these operations are correct). For the server-side costs and the network overhead, we use a mix of the released code combined with reported numbers from their paper. 

We emphasize that Mycelium has a different and a stronger threat model than Colo---Mycelium assumes a single byzantine server, while Colo assumes a set of servers (e.g., 40) of which 20\% can be byzantine.
Thus, Mycelium is not a direct comparison (but the closest in the literature), and we use it to situate Colo's costs, as both Mycelium and Colo operate in strong threat models.

\emparagraph{Queries and datasets.}
We measure and report the overhead for the example queries in Figure~\ref{f:query}. Recall that Colo's overhead depends on the size of the input domain of the query predicate, that is, the length
$len(T)$ of array $T$ which contains a value for every possible input to the predicate. 
Thus, we vary $len(T)$.
For example, if we assume $inf$ has 2 possible values, $tInf$ is any value between 30 and 120~\cite{jing2020household,gudbjartsson2020spread,bi2020epidemiology,adam2020clustering}, and $age$ has 120 possibilities, the queries Q1, Q2, Q4, Q7 have $len(T)=2$, Q3, Q5, Q6 have $len(T)\in [60,240]$, and Q8 has $len(T) = 240$. 
Given these numbers, we vary $len(T)$ between 2 and 1000, to cover our example queries. 

For the number of nodes and edges in the underlying graph, we take inspiration from several public datasets from the Stanford Large Network Dataset Collection (SNAP)~\cite{snapnets} (Figure~\ref{fig:dataset}).
In terms of topology, we vary the maximum degree of nodes between 1 and 100; setting this maximum degree is necessary, 
as each device must communicate with a fixed number of nodes (with itself if it does not have enough neighbors) to hide its topology.

\emparagraph{Testbed \& methodology.} Our testbed is Amazon EC2 and we use machines of type c5.4xlarge to run the devices and the servers. Each such  machine 
has 16 cores, 32 GiB memory, and costs \$0.328 per hour~\cite{aws2023c5}. For experiements, we don't run all devices, e.g., 1M, in one go, which would require thousands of machines. Instead, we sequentially run batches of 1K devices each. In a batch, each device connects to e.g. 50 neighbors. This is acceptable as a device's overhead only depends on its neighborhood. Servers have a fixed-cost (dummy/noise messages; described in $\S~\ref{s:eval:devices}$) for 1M devices, plus incremental costs per-device. We measure and aggregate these separately.

The one-time setup phase cost is not included in our experiments because our focus is the per-query cost. For completeness, we estimate this cost: the server-side cost is in the order of minutes of \cpu time and a few GiBs in network transfers~\cite{zokrates2023zokrates, bowe2017scalable} and the device-side cost is 102~MiB ($\S\ref{s:design:setup}$).

\begin{figure*}[t]
\centering
%\hspace*{-2.5em}
\begin{subfigure}[t]{0.4\textwidth}{
\centering
\includegraphics[width=2.5in]{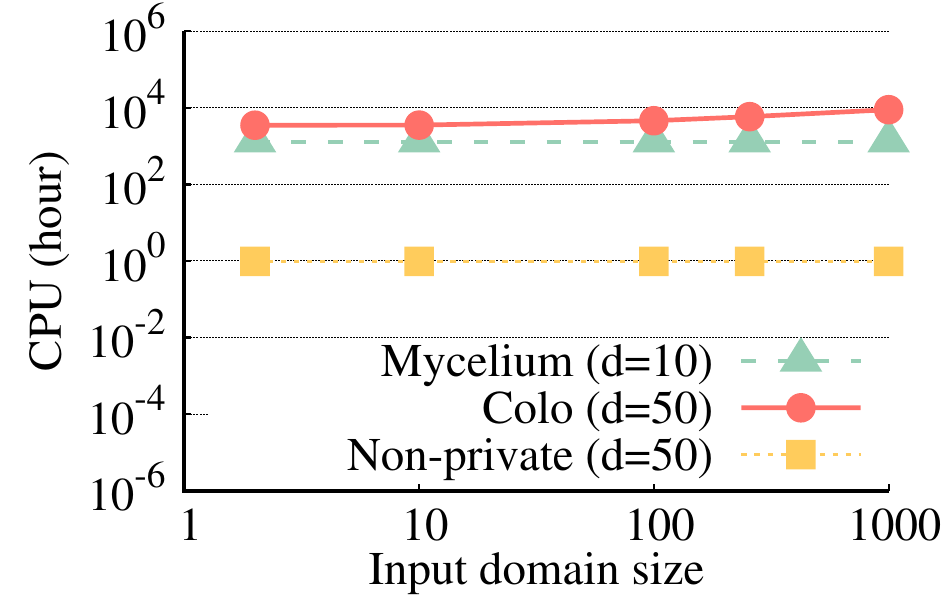}
\caption{Server \cpu}
\label{f:server_cpu_vs_tt}
}
\end{subfigure} %\hspace*{\fill} 
\begin{subfigure}[t]{.4\textwidth}{
\centering
\includegraphics[width=2.5in]{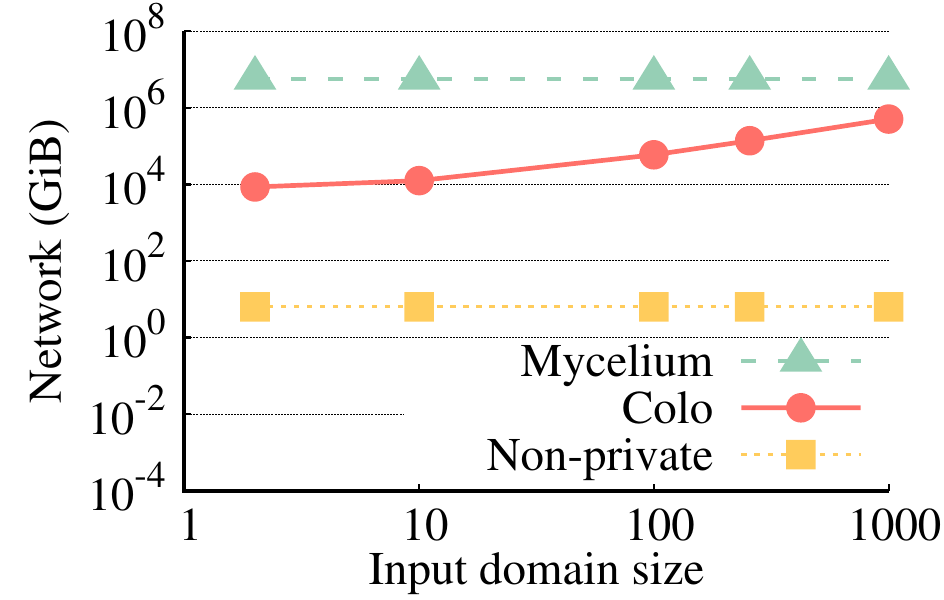}
\caption{Server network}
\label{f:server_net_vs_tt}
} %\hspace*{\fill}
\end{subfigure}
\caption{
Total server-side cost for Colo (across its 40 servers combined) and the baselines with a varying input domain size $len(T)$ of the query predicate. The total number of devices is 1M and the degree of each is 50 (except for Mycelium's \cpu cost where it is 10).} 
\label{f:server_vs_tt}
\label{fig:sever_vs_tt}
\end{figure*}

\subsection{Overhead for different queries}
\label{s:eval:queries}

This section evaluates and compares the overhead of Colo and the baselines for different queries, that is, different sizes of input domain $len(T)$ of the query predicate. More precisely, we fix the query predicate to $F(v_A.input, v_B.input) = v_A.input \cdot v_B.input$, that is, the predicate for query Q1, and set different input sets for $v_A.input$. For instance, $v_A.input \in [0,1]$
fixes $len(T) = 2$, $v_A.input \in [0, 9]$ fixes $len(T) = 10$, and so on.
For these experiments, we fix the number of devices to 1M where each device has at most 50 neighbors.

\emparagraph{Device-side overhead.}
Figure~\ref{f:client_vs_tt} shows per-device \cpu and network overhead as a function of $len(T)$. 
Colo's overhead for $len(T)=2$ (i.e., query Q1, Q2, Q4, and Q7 in Figure~\ref{fig:query}) is 10.27~s and 415~KiB, and increases to 
8.42~min and 4.93~MiB for $len(T)=256$ (this covers all queries in Figure~\ref{fig:query}), and 29.9~min and 18.12~MiB for $len(T) = 1000$.

Colo incurs significantly more overhead than the non-private baseline.
For instance, 
for $len(T) = 256$, 
the baseline's overhead is 0.003~s and 6.36~KiB per device, while 
Colo's overhead is 
$1.68 \cdot 10^{5}$ and $775$ times higher (505.1~s and 4.93~MiB) for the \cpu and the network, respectively.
This is because the non-private baseline does the \changebars{involved}{involve} cryptographic operations, and devices simply share their data with their neighbors, who perform local computations in plaintext. In contrast, in \sys a device's overhead is dominated by the cost
of running the local aggregation protocol (\S\ref{s:design:localaggregation:1}) that involves primitives such as commitments, ZKP, and OT (Figure~\ref{f:localaggregation}).

However, relative to Mycelium, Colo's overhead are significantly lower. For instance, for $len(T) = 256$, 
a device in Mycelium requires 
31, 650~s \cpu time (8.79~h) and 5.73~GiB network transfers, 
while Colo's 505.1~s and 4.93~MiB is $62.6\times$ and $1.16 \cdot 10^{3}\times$ lower.
Mycelium's costs are higher because of various reasons.
First, it performs local aggregation using homomorphic encryption and ZKP. 
For instance, devices multiply ciphertexts containing the inputs of
the neighbors,
and prove that this multiplication is correct using a ZKP (that a malicious device did not manipulate the computation to 
reveal another device's input).
Second, the homomorphic encryption ciphertexts are large (e.g., 4.3~MiB) as Mycelium must choose large parameters for homomorphic encryption to support 
ciphertext-ciphertext multiplications.
This means that the circuit to do the proof is also large, thereby requiring hundreds of seconds of proving time per ciphertext multiplication. 
Third, a Mycelium device 
must 
participate in a verification protocol for global aggregation (besides participating in local aggregation) to check the 
byzantine aggregator's work.
And, fourth, Mycelium also instantiates a mixnet over the devices to hide their topology data, 
again contributing to the device overhead.
In contrast to Mycelium, \sys's cost is dominated by local aggregation only as 
the metatdata-hiding communication is instantiated over the resourceful servers, and the global aggregation is also lightweight (\S\ref{s:overview}, \S\ref{s:design:localaggregation:2}, \S\ref{s:design:globalaggregation}).
Further, the dominant
local aggregation requires
ZKP-friendly commitments and range proofs alongside OT, primitives that have been optimized 
in the literature (\S\ref{s:design:localaggregation:1}). 

Although \sys's overheads are lower, they
increase with $len(T)$ as the number of cryptographic operations in Colo's local aggregation protocol depend linearly on $len(T)$. 
For instance, 
the \cpu time of a Colo device is dominated by the time
to generate the ZKP---that each entry of $T$ is within a range. This cost depends on the number of entries, with the proof for a single entry taking 0.04~s. 
Similarly, a Colo device's network overhead increases approximately linearly with $len(T)$; for instance, a device sends 
one commitment per entry of $T$.
In contrast, the overheads for the baselines
do not depend on $len(T)$ as their computation model
multiplies
$v_A.input \cdot v_B.input$ directly, rather than enumerating all possible outputs. Thus, Colo's overheads will surpass that of Mycelium for a large $len(T)$.

Overall, for Colo's target queries, 
its device-side overheads (in the range of a few seconds to a few minutes in \cpu time, and a few hundred KiBs to a few MiBs in network transfers)
appear to be in the realm of practicality for modern 
mobile devices. 

\emparagraph{Server-side overhead.}
Figure~\ref{f:server_vs_tt} shows the server-side overhead for Colo and the baseline systems. 
Colo's server-side cost
    for $len(T) = 2$ is 88~h \cpu time (on a single core) and 214.78~GiB per server, 
    increases to 147.36~h and 3.46~TiB for $len(T) = 256$, 
    and 224.17~h and 12.79~TiB
    for $len(T) = 1000$.

As with the device-side overhead, Colo's costs 
are significantly higher than the non-private baseline, 
whose server incurs 58.6~min \cpu time and 6.36~GiB network transfers. The baseline server only needs to forward 32-bit messages from devices to other devices; in contrast, Colo's servers are responsible for hiding topology. They not only forward onion-encrypted messages from the devices over $m=28$ hops, but also generate noise messages as part of the Karaoke system to protect against malicious servers (\S\ref{s:design:localaggregation:2}).

Relative to Mycelium, Colo's server-side \cpu is comparable, while the network is significantly lower. 
(For Mycelium's server-side \cpu, since we take numbers from their paper we are only able to report 
for a degree bound of 10, which is what they experiment with.)
For instance, for $len(T)=256$, the server-side cost
for Mycelium is 1304~h and 5737~TiB, while it is 5894~h and 138.4~TiB for \sys's 40 servers combined.
The \cpu cost for Mycelium is dominated by the cost to verify ZKPs (for homomorphic operations) and perform global aggregation (over its byzantine server), while the \cpu cost for \sys is dominated by the time to process onion-encrypted messages. Both overhead turn out to be comparable. Meanwhile, the network overhead due to these dominant operations is lower in Colo. 
Although Colo's server-side costs are substantial, since servers are resourceful (e.g., the \cpu time can be split across many cores), we consider it affordable. We further assess this affordability by calculating the dollar cost of renting the servers.
\begin{figure*}[t]
\centering
\begin{subfigure}[t]{0.4\textwidth}{
\centering
\includegraphics[width=2.5in]{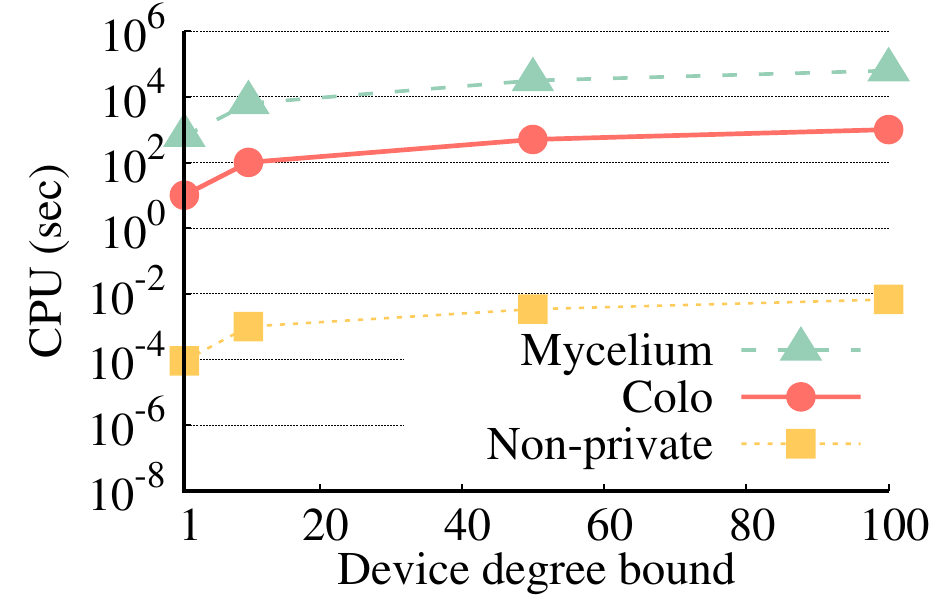}
\caption{Device \cpu}
\label{f:client_cpu_vs_nd}
}
\end{subfigure} %\hspace*{\fill} 
\begin{subfigure}[t]{.4\textwidth}{
\centering
\includegraphics[width=2.5in]{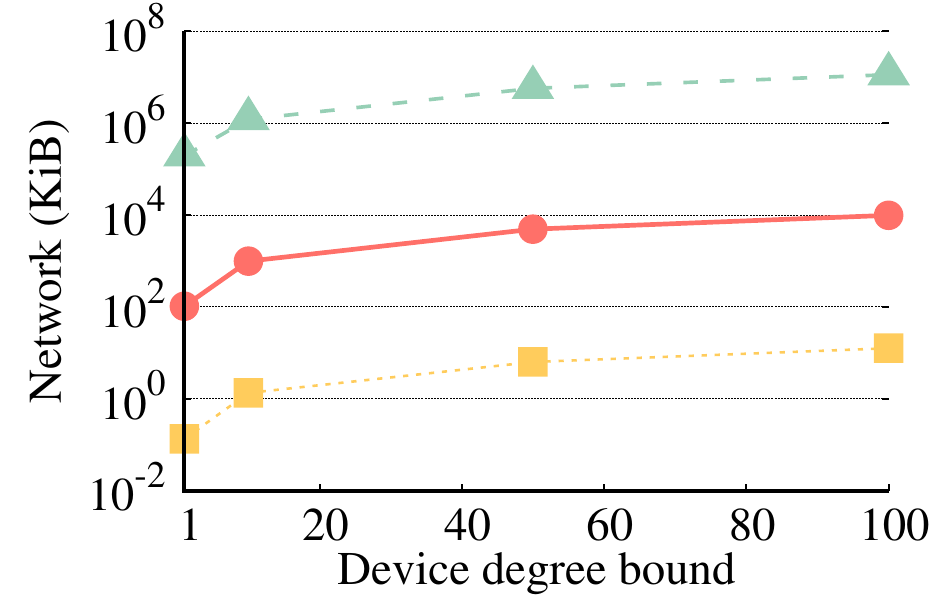}
\caption{Device network}
\label{f:client_net_vs_nd}
}
\end{subfigure}
\caption{
Device-side costs with a varying node degree. These experiments set $N=1M$ devices and
input domain $len(T) = 256$ for the query.} 
\label{f:client_vs_nd}
\label{fig:client_vs_nd}
\end{figure*}

We take both the \cpu and network cost and convert it into a dollar cost per query using the following pricing model.
Each c5.4xlarge machine on AWS
costs \$0.328 per hour and has 16 cores and 32~GiB memory~\cite{aws2023c5}. We assume 
that all hourly cost is due to the \cpu, and compute the \cpu cost per hour to be \$0.0205. For the network cost, we set the cost of transferring one GiB to \$0.01 according to the typical bulk network pricing of cloud providers~\cite{azure2023net,aws2023net}.

Applying this pricing model, \sys's dollar cost per query (over 1M devices) ranges from \$3.95 to \$37.6 per server, 
or \$158 to \$1504 total for the 40 servers, depending on the query. 
In contrast, the non-private baseline costs \$0.08 and Mycelium costs \$57,490 per query.
These figures for Colo are substantial but within the reach of the budget of an entity like CDC. 
For instance, running the cheapest query (which includes the superspreader query) every two weeks
will cost less than $\$4$K annually. 

\subsection{Overhead with the degree of devices}
\label{s:eval:degree}
In this subsection, we evaluate 
how the device-side costs change with the degree bound (the maximum degree of a node). 
Recall that setting a maximum degree is necessary, 
as each device must communicate with a fixed number of nodes (with itself if it does not have enough neighbors) to hide the communication pattern.

Figure~\ref{fig:client_vs_nd} shows the device-side \cpu and network costs for the three systems for 
degree bounds of 1, 10, 50, and 100, for $N = 1$M devices and $len(T) = 256$.
We report only the device-side costs here because 
the server-side costs change with the total number of devices, and increasing the degree bound is equivalent to 
increasing the number of devices from a server's point of view (\S\ref{s:eval:devices}). 

When the degree bound is 10, \sys's devices spend 1.68~min in \cpu and 989.77~KiB in network transfers. This cost increases to 16.8~min \cpu and 9.85~MiB network with a degree bound of 100.
The costs increase and decrease with the degree bound because  each device participates in one instance of secure computation (Figure~\ref{f:localaggregation}) 
per neighbor. 
This linear dependence is present even for the baselines. 
For instance, Mycelium's per-device costs are 1.75~h and 1.16~GiB for a degree bound of 10, and 17.5~h and 11.16~GiB for a degree bound of 100.
Overall, \sys can scale to a significant degree bound (e.g., 10 or 50) with affordable costs for the devices.

\begin{figure*}[t]
\centering
%\hspace*{-2.5em}
\begin{subfigure}[t]{0.33\textwidth}{
\centering
\includegraphics[width=2.25in]{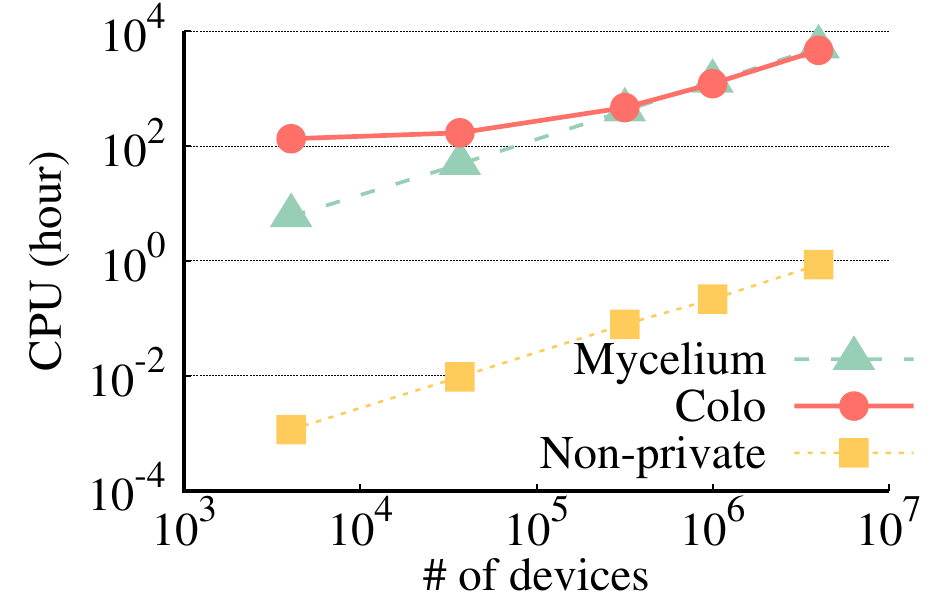}
\caption{Server \cpu}
\label{f:server_cpu}
}
\end{subfigure} %\hspace*{\fill} 
\begin{subfigure}[t]{.33\textwidth}{
\centering
\includegraphics[width=2.25in]{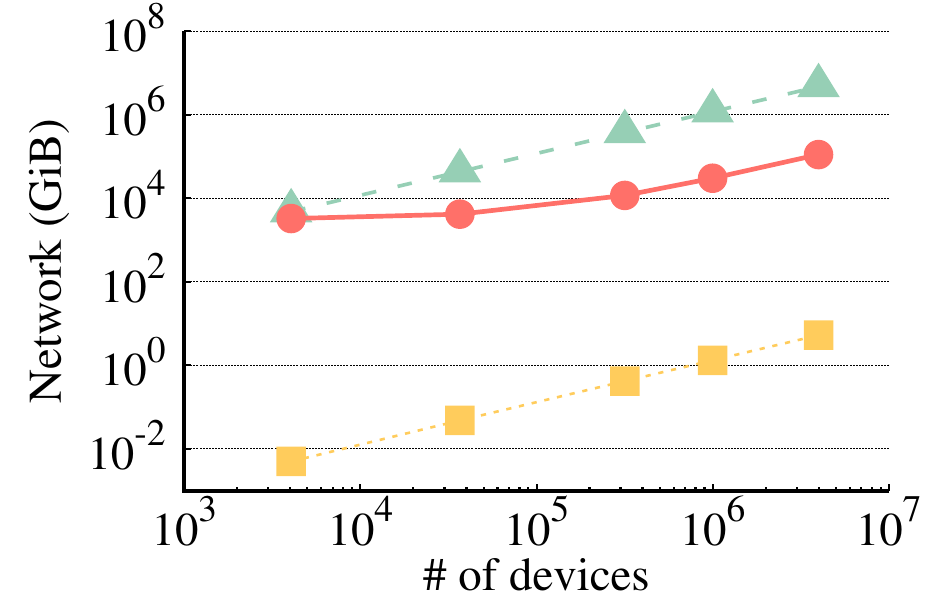}
\caption{Server network}
\label{f:server_net}
} %\hspace*{\fill}
\end{subfigure}
\begin{subfigure}[t]{.33\textwidth}{
\centering
\includegraphics[width=2.25in]{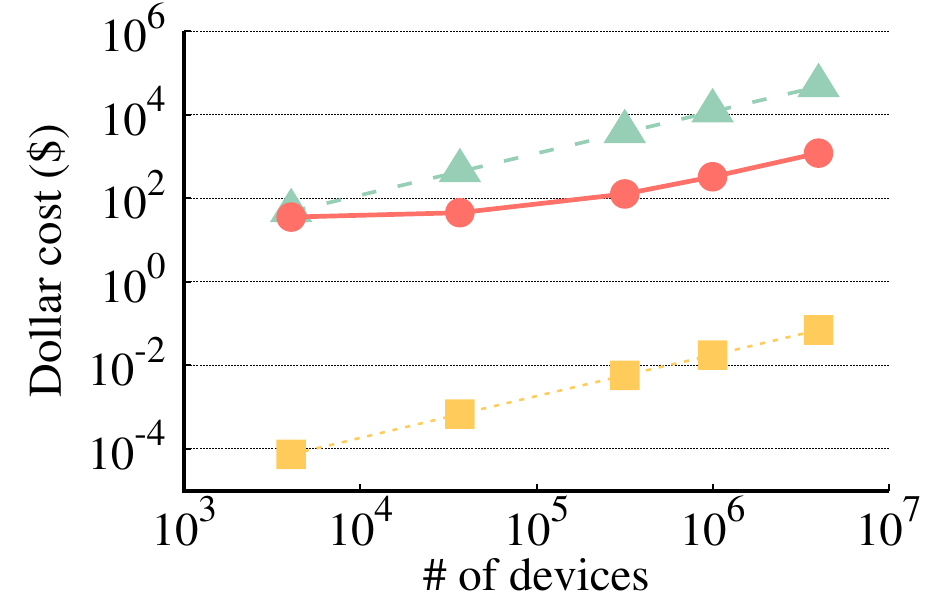}
\caption{Dollar cost}
\label{f:server_dolloar}
} %\hspace*{\fill}
\end{subfigure}
\caption{
Total server-side cost for Colo (across its 40 servers combined) and the baselines with a varying number of devices $N$. 
The degree of each device is 10 and the input domain size of the query predicate is set to $len(T) = 256$.}
\label{f:server_vs_clients}
\label{fig:server_vs_clients}
\end{figure*}

\subsection{Overhead with the number of devices}
\label{s:eval:devices}

This subsection evaluates how the three systems scale with the total number of devices. 
Specifically, we set the number of devices (nodes) from our example datasets (Figure~\ref{f:dataset}), 
that is, $N \in \{4K, 36.6K, 317K, 3.99M\}$. We also consider $N = 1M$.
We fix $len(T) = 256$ and the maximum degree of a node to 10 in alignment with Mycelium's default setting for a better comparison to its server-side \cpu costs (\S\ref{s:eval:queries}). Further, since the device-side costs do not depend on the number of devices, rather just the configuration of their local neighborhood, we focus on the server-side costs.

Figure~\ref{fig:server_vs_clients} shows these costs, that is, 
\cpu time, network transfers, and dollar cost for the servers.
\Sys's cost  is
3.37~h \cpu time, 82.15~GiB network transfers, and 0.89 dollars, per server, for $N=4039$ devices, 
increases to 
 30.74~h \cpu time, 758.09~GiB network transfers, and 8.21 dollars per server for $N=1M$ devices,
and 
further increases to 116.67~h \cpu time, 2.79~TiB network transfers, and 30.29 dollars per server for $N=3.99M$ devices.

The server-side costs for Colo (and the baselines) increases with the number of devices. This is expected as the servers do more work for more devices: send more onion-encrypted messages in the case of \sys, 
route more plaintext messages in the case of the non-private baseline, and operate over more homomorphic encryption ciphertexts (e.g., in global aggregation) for Mycelium.
However, \sys's cost is flatter and higher for a lower $N$. 
This is because even for a small number of devices, \sys's
servers must generate and process several million
noise messages (\S\ref{s:design:localaggregation:2}) 
as required by Karaoke to protect 
against malicious servers dropping traffic to learn access patterns to dead drops. 
For instance, 
for $N=4039$ devices, 
3.25~h (96.4\%) and 79.41~GiB (96.7\%) of the server-side cost in Colo comes from the noise messages.
However, this fixed cost becomes less significant as the number of devices increases, 
thus bringing \sys's costs lower than Mycelium.
For instance, 
\sys incurs 35.6 dollars in total for its 40 servers relative to 48 dollars in Mycelium ($1.34\times$)
for $N=4K$, but 1211.6 dollars versus
47,690 dollars for Mycelium ($39.3\times$) when 
$N$ equals $3.99M$.

Overall, if we consider the dollar cost per server as a key metric for \sys's costs, then \sys appears to scale to several million devices. 
However, scaling it beyond 
these values would require further optimizations and refinements at its servers.

\section{Related work}\label{s:relwork}

\subsection{Graph analytics with a single data owner}
A long line of research 
in graph analytics~\cite{li2023flare, du2022graphshield, xie2014cryptgraph, wang2022pegraph,lai2019graphse2, liu2018enabling, luo2022approximate, mazloom2018secure,mazloom2020secure,wang2022privacy, sharma2018privategraph,wang2022oblivgm, nayak2015graphsc, araki2021secure, wang2023mago}
focuses on outsourcing where
a single data owner outsources computations over its graph to one
untrusted server or a set of non-colluding servers. One common goal for these works is to
hide the graph data from the server(s). For this purpose, they use different cryptographic tools 
or assumptions.

Flare~\cite{li2023flare}, for instance, 
outsources computation to servers with a trusted execution environment (TEE), e.g., Intel SGX. Specifically, it runs 
Spark-like computation inside TEE, which can support the GraphX interface. 
Other systems~\cite{du2022graphshield, xie2014cryptgraph, wang2022pegraph,lai2019graphse2, liu2018enabling, luo2022approximate}
use homomorphic encryption instead of TEEs to run a variety of graph queries on an encrypted graph, including search~\cite{wang2022pegraph,lai2019graphse2} and shortest distance computations~\cite{liu2018enabling, luo2022approximate}.
In the multiple-server setting, a common technique is to use secret sharing and multiparty computation to hide the graph from the servers~\cite{mazloom2018secure,mazloom2020secure,wang2022privacy, sharma2018privategraph,wang2022oblivgm, nayak2015graphsc, araki2021secure}.
For instance, GraphSC~\cite{nayak2015graphsc} assumes two honest-but-curious servers and supports parallel execution of secure computation for a broad class of tasks. 
While GraphSC targets honest-but-curious servers, 
MAGO~\cite{wang2023mago}, which is designed for subgraph counting, assumes three servers where one can be malicious.

The biggest difference between Colo and these existing works is the problem setting. 
While these prior works target a single data owner who outsources computation, 
Colo targets a setting with many data owners (e.g., a million devices). Naturally, in Colo's setting
it is natural to consider the threat of malicious devices. This difference in setting and threat model leads to a vastly different system architecture (\S\ref{s:arch}) and protocol design (\S\ref{s:design}).

\subsection{Federated graph analytics}
Private data federation~\cite{mcsherry2009privacy, narayan2012djoin, 
    bater2018shrinkwrap, han2022scape, bater2017smcql, poddar2021senate}
focuses on a scenario where
a set of data owners hold private relational data and a query coordinator 
orchestrates SQL queries on this data. The goal is to keep the 
sensitive data of the data owners private. Private data federation has two differences with Colo. First, they natively target the relational data model while Colo targets the graph data model. (We include these works here because sometimes graph queries can be expressed in the relational model.) 
Second, these existing works target a scenario with a few data owners (e.g., a few tens) where each
holds a significant partition of the relational data. For instance, 
Senate~\cite{poddar2021senate}
experiments with 16 parties. In contrast, 
Colo targets millions of participants where each has a small amount of data.
Third, while most works in private data federation consider parties to be honest-but-curious, it 
is natural in Colo to consider malicious parties given their number.

DStress~\cite{papadimitriou2017dstress} focuses on graph queries
and a larger number of participants: a few thousand. The key
use case is understanding systemic risk for financial institutions, 
which requires analyzing inter-dependencies across institutions. However, DStress
also assumes honest-but-curious participants as financial institutions 
are heavily regulated and audited, and thus unlikely to be malicious.

Gunther et al.~\cite{gunther2022poster,gunther2022privacy} consider malicious devices alongside a set of semi-honest servers to answer epidemiological queries such as an estimate of the change in the number infections if schools are closed for 14 days.
However, they are limited to secure aggregation across neighbors' data and their protocol does not hide the result of local aggregation.

Mycelium~\cite{roth2021mycelium} is the closest related work to Colo. It supports a broad set of queries, targets
a large number of devices and assumes they can be malicious. 
However, as discussed earlier (\S\ref{s:problem:challenges}) and evaluated empirically (\S\ref{s:eval}), its costs are very high. Colo is a more affordable alternative in a strong threat model, at least for the subset of queries that Colo targets.

\section{Summary}
\label{s:summary}
Privacy-preserving federated graph analytics is an important problem as graphs are natural in many contexts. It specifically is appealing because it keeps 
raw data at the devices (without centralizing) and 
does not release any intermediate computation result except for the final query result. However, the state-of-the-art prior work for this problem is expensive, especially for the devices that have constrained resources. We presented Colo, a new system that operates in a strong threat model while considering malicious devices. Colo addresses the challenge of gaining on efficiency through a new secure computation protocol that allows devices to compute privately and efficiently with their neighbors while hiding node, edge, and topology data (\S\ref{s:design:localaggregation:1}, \S\ref{s:design:localaggregation:1}, \S\ref{s:overview}). 
The per-device overhead in Colo is a few minutes of \cpu and a few MiB in network transfers, while the server's overhead ranges from several dollars to tens of dollars per server, per query (\S\ref{s:eval}).
Our conclusion is that Colo brings privacy-preserving federated graph analytics into the realm of practicality 
for a certain class of queries.

\clearpage

\goodcitationsize
\begin{flushleft}
\setlength{\parskip}{0pt}
\setlength{\itemsep}{0pt}
\bibliographystyle{abbrv}
\bibliography{conferences-long-with-abbr2,paper}
\end{flushleft}
\clearpage

\end{document}